\documentclass{article}
\usepackage{fullpage}
\usepackage{amsmath}
\usepackage{url}
\usepackage{verbatim}
\usepackage[square, comma, sort&compress]{natbib}
\begin{document}
\large

\title{Haplotype-based variant detection from short-read sequencing}
\author{Erik Garrison and Gabor Marth}

\maketitle

\begin{abstract}
The direct detection of haplotypes from short-read DNA sequencing data
requires changes to existing small-variant detection methods.  Here,
we develop a Bayesian statistical framework which is capable of
modeling multiallelic loci in sets of individuals with non-uniform
copy number.  We then describe our implementation of this framework in
a haplotype-based variant detector, FreeBayes.
\end{abstract}

\section{Motivation}

\begin{comment}
Fully understanding the functional impact of genetic variation
requires a determination of whether nearby variants are found on the
same parental chromosome, also known as “in phase.”  A pair of
variants discovered in a single exon, transcription factor binding
site, or splice junction may have dramatically different effects if
they are in phase than if they are found as separately segregating
alleles.  Similarly, the local haplotype structure surrounding a
variant of interest is essential when designing probe or primer-based
genotyping assays, and the failure to incorporate nearby linked
variants into assay design is a known cause of genotyping failure for
such methods.
\end{comment}

\begin{comment}
For example, two frameshift indels in the same exon may
induce an overall non-frameshift effect on the exon if they lie on the
same haplotype.  Analogously, a pair of SNPs overlapping a protein
binding site may generate a combined effeoct on binding that is
substantially greater if they are in phase than if they are not.
\end{comment}

\begin{comment}
Current high-performance sequence variant detection methods apply a
single-locus-based detection model to short-read sequencing alignment
data.  This approach to primary variant detection requires the
reconstruction of phasing information \emph{post hoc} using
computationally expensive statistical approaches that depend on
information combined across many samples from the same population
\citep{browning2007, mach2010, delaneau2012, howie2011}.
\end{comment}

While \emph{statistical phasing} approaches are necessary for the
determination of large-scale haplotype structure \citep{browning2007,
  mach2010, delaneau2012, howie2011}, sequencing traces provide
short-range phasing information that may be employed directly in
primary variant detection to establish phase between proximal alleles.
Present read lengths and error rates limit this \emph{physical
  phasing} approach to variants clustered within tens of bases, but as
the cost of obtaining long sequencing traces decreases
\citep{branton2008, clarke2009}, physical phasing methods will enable
the determination of larger haplotype structure directly using only
sequence information from a single sample.

Haplotype-based variant detection methods, in which short haplotypes
are read directly from sequencing traces, offer a number of benefits
over methods which operate on a single position at a time.
Haplotype-based methods ensure semantic consistency among described
variants by simultaneously evaluating all classes of alleles in the
same context.  The use of locally-phased genotype data can lower the
computational burden of genotype imputation by reducing the possible
space of haplotypes which must be considered.  Locally phased
genotypes can be used to improve genotyping accuracy in the context of
rare variations that can be difficult to impute due to sparse linkage
information.  Similarly, they can assist in the design of genotyping
assays, which can fail in the context of undescribed variation at the
assayed locus.  Provided sequencing errors are independent, the use of
longer haplotypes in variant detection can improve detection by
increasing the signal to noise ratio of the genotype likelihood space
that is used in analysis.  This follows from the fact that the space
of possible erroneous haplotypes expands dramatically with haplotype
length, while the space of true variation remains constant, with the
number of true alleles less than or equal to the ploidy of the sample
at a given locus.

The direct detection of haplotypes from alignment data presents
several challenges to existing variant detection methods.  As the
length of a haplotype increases, so does the number of possible
alleles within the haplotype, and thus methods designed to detect
genetic variation over haplotypes in a unified context must be able to
model multiallelism.  However, most variant detection methods
establish estimates of the likelihood of polymorphism at a given loci
using statistical models which assume biallelism
\citep{li2011stats,marth99} and uniform, typically diploid, copy
number \citep{gatk2011}.  Moreover, improper modeling of copy number
impedes the accurate detection of small variants on sex chromosomes,
in polyploid organisms, or in locations with known copy-number
variations, where called alleles, genotypes, and likelihoods should
reflect local copy number and global ploidy.

To enable the application of population-level inference methods to the
detection of haplotypes, we generalize the Bayesian statistical method
described by \citet{marth99} to allow multiallelic loci and
non-uniform copy number across the samples under consideration.  We
have implemented this model in FreeBayes \citep{freebayesgit}.

\section{Generalizing variant detection to multiallelic loci and non-uniform copy number}
\label{sec:model}

\subsection{Definitions}

At a given genetic locus we have $n$ samples drawn from a population,
each of which has a copy number or multiplicity of $m$ within the
locus.  We denote the number of copies of the locus present within our
set of samples as $M = \sum_{i=1}^n m_i$.  Among these $M$ copies we
have $K$ distinct alleles, $b_1,\ldots,b_K$ with allele frequencies
$f_1,\ldots,f_K$.  Each individual has an unphased genotype $G_i$
comprised of $k_i$ distinct alleles $b_{i_1},\ldots,b_{k_i}$ and
corresponding genotype allele frequencies
$f_{i_1},\ldots,f_{i_{k_i}}$, which may be equivalently expressed as a
multiset of alleles $B_i : | B_i | = m_i$.  For the purposes of our
analysis, we assume that we cannot accurately discern phasing
information outside of the haplotype detection window, so our $G_i$
are unordered and all $G_i$ containing equivalent alleles and
frequencies are regarded as equivalent.  Assume a set of $s_i$
sequencing observations $r_{i_1},\ldots,r_{i_{s_i}} = R_i$ over each
sample in our set of $n$ samples such that there are $ \sum_{i=1}^n
s_i$ reads at the genetic locus under analysis.  $Q_i$ denotes the
mapping quality, or probability that the read $r_i$ is mis-mapped
against the reference.

\subsection{A Bayesian approach}
\label{sec:modeloverview}

To genotype the samples at a specific locus, we could simply apply a
Bayesian statistic relating $P(G_i|R_i)$ to the likelihood of
sequencing errors in our reads and the prior likelihood of specific
genotypes.  However, this maximum-likelihood approach limits our
ability to incorporate information from other individuals in the
population under analysis, which can improve detection power.

Given a set of genotypes $G_1,\ldots,G_n$ and observations
observations $R_1,\ldots,R_n$ for all individuals at the current
genetic locus, we can use Bayes' theorem to related the probability of
a specific combination of genotypes to both the quality of sequencing
observations and \emph{a priori} expectations about the distribution
of alleles within a set of individuals sampled from the same population:

\begin{equation}
P(G_1,\ldots,G_n | R_1,\ldots,R_n) = { P(G_1,\ldots,G_n) P(R_1,\ldots,R_n | G_1,\ldots,G_n) \over P(R_1,\ldots,R_n)} \\
\end{equation}

\begin{equation}
\label{eq:bayesian}
P(G_1,\ldots,G_n | R_1,\ldots,R_n)  = { P(G_1,\ldots,G_n) \prod_{i=1}^n P(R_i|G_i) \over 
\sum_{\forall{G_1,\ldots,G_n}} ( P(G_1,\ldots,G_n) \prod_{i=1}^n P(R_i|G_i) ) }
\end{equation}

Under this decomposition, $P(R_1,\ldots,R_n|G_1,\ldots,G_n) =
\prod_{i=1}^n P(R_i|G_i)$ represents the likelihood that our
observations match a given genotype combination (our data likelihood),
and $P(G_1,\ldots,G_n)$ represents the prior likelihood of observing a
specific genotype combination.  We estimate the data likelihood as the
joint probability that the observations for a specific individual
support a given genotype.  We use a neutral model of allele diffusion
conditioned on an estimated population mutation rate to estimate the
prior probability of sampling a given collection of genotypes.

Except for situations with small numbers of samples and alleles, we
avoid the explicit evaluation of the posterior distribution as implied
by (\ref{eq:bayesian}), instead using a number of optimizations to make the algorithm
tractable to apply to very large datasets (see section \ref{sec:genotyping}).

\subsection{Estimating the probability of a sample genotype given sequencing observations, $P(R_i|G_i)$}

Given a set of reads $R_i = r_{i_1},\ldots,r_{i_{s_i}}$ of a sample at
a given locus, we can extract a set of $k_i$ observed alleles $B'_i =
b'_1,\ldots,b'_{k_i}$ which encapsulate the potential set of
represented variants at the locus in the given sample, including
erroneous observations.  Each of these observed alleles $b'_l$ has a
frequency $o'_f$ within the observations of the individual sample $:
\sum_{j=1}^{k_i} o'_j = s_i$ and corresponds to a true allele $b_l$.

If we had perfect observations of a locus, $P(R_i|G_i)$ for any
individual would approximate the probability of sampling observations
$R_i$ out of $G_i$ with replacement.  This probability is given by the
multinomial distribution in $s_i$ over the probability $P(b_l)$ of
obtaining each allele from the given genotype, which is ${f_{i_j}
  \over m_i}$ for each allele frequency in the frequencies which
define $G_i$, $f_{i_1},\ldots,f_{i_{k_i}}$.

\begin{equation}
P(R_i | G_i)
\approx
P(B'_i | G_i) = 
{ s_i! \over { 
\prod_{j=1}^{k_i} o'_j !
} }
\prod_{j=1}^{k_i} { \left({f_{i_j} \over m_i}\right)^{o'_j} }
\end{equation}

Our observations are not perfect, and thus we must account for the
probability of errors in the reads.  We can use the per-base quality
scores provided by sequencing systems to develop the probability that
an observed allele is drawn from an underlying true allele,
$P(B'_i|R_i)$.  We assume a mapping between sequencing quality scores
and allele qualities such that each observed allele $b'_l$ has a
corresponding quality $q_l$ which approximates the probability that
the observed allele is incorrect.

Furthermore, we must sum $P(R_i|G_i)$ for all possible $R_i$
combinations $\forall(R_i \in G_i : | R_i | = k_i)$ drawn from our
genotype to obtain the joint probability of $R_i$ given $G_i$, as each
$\prod_{l=1}^{s_i} { P(b'_l | b_l) }$ only accounts for the marginal
probability of the a specific $R_i$ given $B'_i$.

This extends $P(R_i|G_i)$ as follows:

\begin{equation}
P(R_i | G_i) =  
\sum_{\forall(R_i \in G_i)} \left(
{ s_i! \over { 
\prod_{j=1}^{k_i} o'_j !
} }
\prod_{j=1}^{k_i} { \left({f_{i_j} \over m_i}\right)^{o'_j} }
\prod_{l=1}^{s_i} { P(b'_l | b_l) }
\right)
\end{equation}

In summary, the probability of obtaining a set of reads given an
underlying genotype is proportional to the probability of sampling the
set of observations from the underlying genotype, scaled by the
probability that our reads are correct.  As $q_l$ approximates the
probability that a specific $b_l$ is incorrect, $P(b'_l|b_l) = 1 -
q_l$ when $b_l \in G_i$ and $P(b'_l|b_l) = q_l$ when $b_l \notin G_i$.

\subsection{Priors for unphased genotype combinations, $P(G_1,\ldots,G_n)$}

\subsubsection{Decomposition of prior probability of genotype combination}

Let $G_1,\ldots,G_n$ denote the set of genotypes at the locus and
$f_1,\ldots,f_k$ denote the set of allele frequencies which
corresponds to these genotypes.  We estimate the prior likelihood of
observing a specific combination of genotypes within a given locus by
decomposition into resolvable terms:

\begin{equation}
P(G_1,\ldots,G_n) = P(G_1,\ldots,G_n \cap f_1,\ldots,f_k)
\end{equation}

The probability of a given genotype combination is equivalent to the
intersection of that probability and the probability of the
corresponding set of allele frequencies.  This identity follows from
the fact that the allele frequencies are derived from the set of
genotypes and we always will have the same $f_1,\ldots,f_k$ for any
equivalent $G_1,\ldots,G_n$.

Following Bayes' Rule, this identity further decomposes to:

\begin{equation}
P(G_1,\ldots,G_n \cap f_1,\ldots,f_k) = P(G_1,\ldots,G_n | f_1,\ldots,f_k) P(f_1,\ldots,f_k)
\end{equation}

We now can estimate the prior probability of $G_1,\ldots,G_n$ in terms
of the genotype combination sampling probability, $P(G_1,\ldots,G_n |
f_1,\ldots,f_k)$, and the probability of observing a given allele
frequency in our population, $P(f_1,\ldots,f_k)$.

\subsubsection{Genotype combination sampling probability $P(G_1,\ldots,G_n | f_1,\ldots,f_k)$}

The multinomial coefficient ${M \choose f_1,\ldots,f_k }$ gives the
number of ways which a set of alleles with frequencies
$f_1,\ldots,f_k$ may be distributed among $M$ copies of a locus.  For
phased genotypes $\hat{G_i}$ the probability of sampling a specific
$\hat{G_1},\ldots,\hat{G_n}$ given allele frequencies $f_1,\ldots,f_k$
is thus provided by the inverse of this term:

\begin{equation}
\label{eq:phasedsampling}
P(\hat{G_1},\ldots,\hat{G_n} | f_1,\ldots,f_k) =
{M \choose
  f_1,\ldots,f_k }^{-1}
\end{equation}

However, our model is limited to unphased genotypes because our
primary data only allows phasing within a limited context.
Consequently, we must adjust (\ref{eq:phasedsampling}) to reflect the
number of phased genotypes which correspond to the unphased genotyping
$G_1,\ldots,G_n$.  Each unphased genotype corresponds to as many
phased genotypes as there are permutations of the alleles in $G_i$.
Thus, for a given unphased genotyping $G_1,\ldots,G_n$, there are
$\prod_{i=1}^n { m_i \choose f_{i_1}, \ldots, f_{i_{k_i}}}$ phased
genotypings.

In conjunction, these two terms provide the probability of sampling a
particular unphased genotype combination given a set of allele
frequencies:

\begin{equation}
\label{eq:unphasedsampling}
P(G_1,\ldots,G_n | f_1,\ldots,f_k) =
{ M \choose f_1,\ldots,f_k }^{-1}
\prod_{i=1}^n { m_i \choose f_{i_1}, \ldots, f_{i_{k_i}}}
 = 
\frac{1}{M!}
\prod_{l=1}^k f_l! 
\prod_{i=1}^n \frac{m_i!}{\prod_{j=1}^{k_i} f_{i_j}!}
\end{equation}

In the case of a fully diploid population, the product of all possible
multiset permutations of all genotypes reduces to $2^h$, where $h$ is
the number of heterozygous genotypes, simplifying
(\ref{eq:unphasedsampling}) to:

\begin{equation}
P(G_1,\ldots,G_n | f_1,\ldots,f_k) =
2^h
{ M \choose f_1,\ldots,f_k }^{-1}
\end{equation}

\subsubsection{Derivation of $P(f_1,\ldots,f_k)$ by Ewens' sampling formula}

Provided our sample size $n$ is small relative to the population which
it samples, and the population is in equilibrium under mutation and
genetic drift, the probability of observing a given set of allele
frequencies at a locus is given by Ewens' sampling formula
\citep{ewens72}.  Ewens' sampling formula is based on an infinite
alleles coalescent model, and relates the probability of observing a
given set of allele frequencies to the number of sampled chromosomes
at the locus ($M$) and the population mutation rate $\theta$.

The application of Ewens' formula to our context is straightforward.
Let $a_f$ be the number of alleles among $b_1,\ldots,b_k$ whose allele
frequency within our set of samples is $f$.  We can thus transform our
set of frequencies $f_1,\ldots,f_k$ into a set of non-negative
frequency counts $a_1,\ldots,a_M : \sum_{f=1}^M{fa_f} = M$.  As many
$f_1,\ldots,f_k$ can map to the same $a_1,\ldots,a_M$, this
transformation is not invertible, but it is unique from
$a_1,\ldots,a_M$ to $f_1,\ldots,f_k$.

Having transformed a set of frequencies over alleles to a set of
frequency counts over frequencies, we can now use Ewens' sampling
formula to approximate $P(f_1,\ldots,f_k)$ given $\theta$:

\begin{comment}
\begin{equation}
P(f_1,\ldots,f_k) = P(a_1,\ldots,a_M) = {M! \over \theta(\theta+1)\cdots(\theta+M-1)}\prod_{j=1}^M{\theta^{a_j} \over j^{a_j} a_j!}
\end{equation}
\end{comment}

\begin{equation}
P(f_1,\ldots,f_k) =
P(a_1,\ldots,a_M) = 
{M! \over \theta \prod_{z=1}^{M-1}(\theta+z)}
\prod_{j=1}^M{\theta^{a_j} \over j^{a_j} a_j!}
\end{equation}

In the bi-allelic case in which our set of samples has two alleles
with frequencies $f_1$ and $f_2$ such that $f_1 + f_2 = M$:

\begin{equation}
P(a_{f_1} = 1, a_{f_2} = 1) = 
{M! \over \prod_{z=1}^{M-1}(\theta+z)}
{\theta \over f_1 f_2}
\end{equation}

While in the monomorphic case, where only a single allele is
represented at this locus in our population, this term reduces to:

\begin{equation}
P(a_M = 1) = 
{(M-1)! \over \prod_{z=1}^{M-1}(\theta+z)}
\end{equation}

In this case, $P(f_1,\ldots,f_k) = 1 - \theta$ when $M = 2$.  This is
sensible as $\theta$ represents the population mutation rate, which
can be estimated from the pairwise heterozygosity rate of any two
chromosomes in the population \citep{watterson1975, tajima1983}.

\section{Direct detection of phase from short-read sequencing}

By modeling multiallelic loci, this Bayesian statistical framework
provides the foundation for the direct detection of longer, multi-base
alleles from sequence alignments.  In this section we describe our
implementation of a haplotype-based variant detection method based on
this model.

Our method assembles haplotype observations over minimal,
dynamically-determined, reference-relative windows which contain
multiple segregating alleles.  To be used in the analysis, haplotype
observations must be derived from aligned reads which are anchored by
reference-matching sequence at both ends of the detection window.
These haplotype observations have derived quality estimations which
allow their incorporation into the general statistical model described
in section \ref{sec:model}.  We then employ a gradient ascent method
to determine the maximum \emph{a posteriori} estimate of a mutual
genotyping over all samples under analysis and establish an estimate
of the probability that the loci is polymorphic.

\subsection{Parsing haplotype observations from sequencing data}
\label{sec:parsing}

In order to establish a range of sequence in which multiple
polymorphisms segregate in the population under analysis, it is
necessary to first determine potentially polymorphic windows in order
to bound the analysis.  This determination is complicated by the fact
that a strict windowing can inappropriately break clusters of alleles
into multiple variant calls.  We employ a dynamic windowing approach
that is driven by the observation of multiple proximal
reference-relative variations (SNPs and indels) in input alignments.

Where reference-relative variations are separated by less than a
configurable number of non-polymorphic bases in an aligned sequence
trace, our method combines them into a single haplotype allele
observation, $H_i$.  The observational quality of these haplotype
alleles is given as $\min ( q_l \, \forall \, b'_i \in H_i , \, Q_i)$,
or the minimum of the supporting read's mapping quality and the
minimum base quality of the haplotype's component variant allele
observations.

\subsection{Determining a window over which to assemble haplotype observations}

At each position in the reference, we collect allele observations
derived from alignments as described in \ref{sec:parsing}.  To improve
performance, we apply a set of input filters to exclude alleles from
the analysis which are highly unlikely to be true.  These filters
require a minimum number of alternate observations and a minimum sum
of base qualities in a single sample in order to incorporate a
putative allele and its observations into the analysis.

We then determine a haplotype length over which to genotype samples by
a bounded iterative process.  We first determine the allele passing
the input filters which is longest relative to the reference.  For
instance, a longer allele could be a multi-base indel or a composite
haplotype observation flanked by SNPs.  Then, we parse haplotype
observations from all the alignments which fully overlap this window,
finding the rightmost end of the longest haplotype allele which begins
within the window.  This rightmost position is used to update the
haplotype window length, and a new set of haplotype observations are
assembled from the reads fully overlapping the new window.  This
process repeats until the rightmost end of the window is not partially
overlapped by any haplotype observations which pass the input filters.
This method will converge given reads have finite length and the only
reads which fully overlap the detection window are used in the
analysis.

\subsection{Detection and genotyping of local haplotypes}
\label{sec:genotyping}

Once a window for analysis has been determined, we parse all
fully-overlapping reads into haplotype observations which are anchored
at the boundaries of the window.  Given these sets of sequencing
observations $r_{i_1},\ldots,r_{i_{s_i}} = R_i$ and data likelihoods
$P(R_i|G_i)$ for each sample and possible genotype derived from the
putative alleles, we then determine the probability of polymorphism at
the locus given the Bayesian model described in section
\ref{sec:model}.

To establish a maximum \emph{a posteriori} estimate of the genotype
for each sample, we employ a convergent gradient ascent approach to
the posterior probability distribution over the mutual genotyping
across all samples under our Bayesian model.  This process begins at
the genotyping across all samples $G_1,\ldots,G_n$ where each sample's
genotype is the maximum-likelihood genotype given the data likelihood
$P(R_i|G_i)$:

\begin{equation}
G_1,\ldots,G_n =
\underset{G_i}{\operatorname{argmax}} \; P(R_i|G_i)
%:=  \{ G_i | \forall G : P(R_i|G_i) >= P(R_i|G) \}
\end{equation}

The posterior search then attempts to find a genotyping
$G_1,\ldots,G_n$ in the local space of genotypings which has higher
posterior probability under the model than this initial genotyping.
In practice, this step is done by searching through all genotypings in
which a single sample has up to the $N$th best genotype when ranked by
$P(R_i|G_i)$, and $N$ is a small number (e.g. 2).  This search starts
with some set of genotypes $G_1,\ldots,G_n = \{G\}$ and attempts to
find a genotyping $\{G\}'$ such that:

\begin{equation}
P(\{G\}'|R_1,\ldots,R_n) > P(\{G\}|R_1,\ldots,R_n)
\end{equation}

$\{G\}'$ is then used as a basis for the next update step.  This
search iterates until convergence, but in practice must be bounded at
a fixed number of steps in order to ensure optimal performance.  As
the quality of input data increases in coverage and confidence, this
search will converge more quickly because the maximum-likelihood
estimate will lie closer to the maximum \emph{a posteriori} estimate
under the model.

This method incorporates a basic form of genotype imputation into the
detection method, which in practice improves the quality of raw
genotypes produced in primary allele detection and genotyping relative
to methods which only utilize a maximum-likelihood method to determine
genotypes.  Furthermore, this method allows for the determination of
marginal genotype likelihoods via the marginalization of assigned
genotypes for each sample over the posterior probability distribution.

\subsection{Probability of polymorphism}

Provided a maximum \emph{a posteriori} estimate of the genotyping of
all the individuals in our sample, we might like establish an estimate
of the quality of the genotyping.  For this, we can use the
probability that the locus is polymorphic, which means that the number
of distinct alleles at the locus, $K$, is greater than 1.  While in
practice the space of possible genotypings is too great to integrate
over, it is possible to derive the probability that the loci is
polymorphic in our samples by summing across the monomorphic cases:

\begin{equation}
\label{eq:probpoly}
P(K > 1 | R_1,\ldots,R_n)
=
1 - P(K = 1 | R_1,\ldots,R_n)
%=
%1 - \sum_{\forall(G_i,\ldots,G_n : K = 1)} P(G_i,\ldots,G_n|R_i,\ldots,R_n)
\end{equation}

Equation (\ref{eq:probpoly}) thus provides the probability of
polymorphism at the site, which is provided as a quality estimate for
each evaluated locus in the output of FreeBayes.

\subsection{Marginal likelihoods of individual genotypes}

Similarly, we can establish a quality estimate for a single genotype
by summing over the marginal probability of that specific genotype and
sample combination under the model.  The marginal probability of a
given genotype is thus:

\begin{equation}
\label{eq:marginals}
P(G_j|R_i,\ldots,R_n)
=
\sum_{\forall(\{G\} : G_j \in \{G\})}
P(\{G\}|R_i,\ldots,R_n)
\end{equation}

In implementation, the development of this term is more complex, as we
must sample enough genotypings from the posterior in order to obtain
well-normalized marginal likelihoods.  In practice, we marginalize
from the local space of genotypings in which each individual genotype
is no more than a small number of steps in one sample from the maximum
\emph{a posteriori} estimate of $G_i,\ldots,G_n$.  This space is
similar to that used during the posterior search described in section
\ref{sec:genotyping}.  We apply (\ref{eq:marginals}) to it to estimate
marginal genotype likelihoods for the most likely individual
genotypes, which are provided for each sample at each site in the
output of our implementation.

\bibliography{references}{}
\bibliographystyle{plainnat}

\end{document}